\documentstyle[12pt]{article}

\begin{document}

\newcommand{\square}{\vrule height 1.5ex width 1.2ex depth -.1ex }


\newcommand{\II}{\leavevmode\hbox{\small1\kern-3.8pt\normalsize1}}

\newcommand{\inbar}{\,\vrule height1.5ex width.4pt depth0pt}
\newcommand{\CC}{\relax{\hbox{$\inbar\kern-.3em{\rm C}$}}}
\newcommand{\RR}{{\rm I\! R}}
\newcommand{\NN}{{\rm I\! N}}
\newcommand{\ZZ}{\hbox{\sansf Z\kern-0.4em Z}}


\newcommand{\CoinfM}{C_0^\infty(M)}
\newcommand{\CoinfN}{C_0^\infty(N)}
\newcommand{\Coinfd}{C_0^\infty(\RR^d\backslash\{ 0\})}
\newcommand{\Coinf}[1]{C_0^\infty(\RR^{#1}\backslash\{ 0\})}
\newcommand{\Coin}{C_0^\infty(0,\infty)}


\newtheorem{Thm}{Theorem}[section]
\newtheorem{Def}[Thm]{Definition}
\newtheorem{Lem}[Thm]{Lemma}
\newtheorem{Prop}[Thm]{Proposition}
\newtheorem{Cor}[Thm]{Corollary}


\renewcommand{\theequation}{\thesection.\arabic{equation}}
\newcommand{\sect}[1]{\section{#1}\setcounter{equation}{0}}


\newcommand{\AAA}{{\cal A}}
\newcommand{\DD}{{\cal D}}
\newcommand{\OO}{{\cal O}}


\newcommand{\xvec}{{\bf x}}
\newcommand{\rvec}{{\bf \hat{r}}}


\font\sansf=cmss12
\newcommand{\Pt}{\hbox{\sansf P}}
\newcommand{\Tt}{\hbox{\sansf T}}


\newcommand{\Dal}{\fbox{\phantom{${\scriptstyle *}$}}}
\newcommand{\vth}{\vartheta}
\newcommand{\vep}{\varepsilon}

\begin{titlepage}
\renewcommand{\thefootnote}{\fnsymbol{footnote}}

\rightline{BUTP-95/31}
\vspace{0.1in}
\LARGE
\center{Quantum Field Theory on Certain\\ Non-Globally Hyperbolic
Spacetimes}
\Large
\vspace{0.2in}
\center{C.J. Fewster\footnote{E-mail address: fewster@butp.unibe.ch}
and A. Higuchi\footnote{E-mail address: higuchi@butp.unibe.ch}}
\vspace{0.2in}
\large
\center{\em Institut f\"{u}r theoretische Physik, Universit\"{a}t Bern \\
Sidlerstrasse 5, CH-3012 Bern, Switzerland}
\vspace{0.2in}
\center{\today}

\vspace{0.2in}

\begin{abstract}

We study real linear scalar field theory on two simple
non-globally hyperbolic spacetimes containing closed timelike curves
within the framework proposed by Kay for algebraic
quantum field theory on non-globally hyperbolic spacetimes. In this
context, a spacetime $(M,g)$ is said to be `F-quantum compatible'
with a field theory if it admits a $*$-algebra of local observables
for that theory which satisfies a locality condition known as
`F-locality'. Kay's proposal is that, in formulating algebraic
quantum field theory on $(M,g)$, F-locality should be imposed as
a necessary condition on the $*$-algebra of observables.

The spacetimes studied are the 2- and 4-dimensional spacelike
cylinders (Mink\-owski space quotiented by a timelike translation).
Kay has shown that the 4-dimensional spacelike cylinder is F-quantum
compatible with massless fields. We prove that it is also F-quantum
compatible with massive fields and prove the F-quantum compatibility
of the 2-dimensional spacelike cylinder with both massive and
massless fields. In each case, F-quantum compatibility is proved
by constructing a suitable F-local algebra.

\vspace{0.3truecm}
{\noindent {\bf PACS Numbers} 04.62.+v}
\end{abstract}

\setcounter{footnote}{0}
\renewcommand{\thefootnote}{\arabic{footnote}}
\end{titlepage}

\sect{Introduction}

The theory of quantum fields in curved spacetime has been developed
for the most part in the context of globally hyperbolic Lorentzian
spacetimes. These spacetimes are distinguished by the fact that they
contain Cauchy surfaces\footnote{Following \cite{Kay}, we define a
Cauchy surface in a Lorentzian spacetime to be a smooth spacelike
surface intersected precisely once by each inextendible causal
(time-like or null) curve contained in the spacetime. A spacetime is
said to be globally hyperbolic if and only if it contains a Cauchy
surface.}, which entails the global existence and uniqueness of
certain fundamental solutions to the Klein-Gordon and Dirac equations
(in particular the advanced-minus-retarded fundamental solution) which
play a key r\^ole in the quantisation of these theories.

There has, however, been recent interest in quantum field theory on
spacetimes containing closed timelike curves (CTCs), which do not
possess Cauchy surfaces and for which the usual methodology must be
modified.  There are various reasons for this interest. For example,
because such spacetimes are a `source of tension' between quantum
theory and general relativity, one would expect that they might be
associated with or forbidden by non-trivial quantum gravitational
effects (cf.  Hawking's Chronology Protection Conjecture
\cite{Hawk}). Even disregarding back-reaction, the formulation of
quantum field theory in fixed background spacetimes containing CTCs
raises many interesting conceptual problems stemming from the lack of
a global Cauchy surface. As an example, we mention the fact that the
evolution of interacting quantum fields from the far past of an
isolated region of CTCs to its far future turns out to be
non-unitary~\cite{Boul,FPS1,Pol}, which has led to a debate on how
this can be interpreted \cite{Hawk2} or alternatively, how unitarity
may be restored \cite{Arley,FW}.

In the presence of such conceptual problems, there are clearly
advantages in adopting a mathematically
rigorous approach to quantum field theory in curved spacetime,
such as that offered by the algebraic approach, even though this
restricts us (at least for the moment) to the study of linear
(non-self-interacting) field theories. The algebraic approach
has also been
developed largely in the globally hyperbolic case (see~\cite{Kay,KW}
for detailed reviews). Recently, however, Kay~\cite{Kay} and
Yurtsever~\cite{Yurt}
have independently proposed suitable generalisations of this approach
to the
non-globally hyperbolic case. Here, we will work within the framework
developed by Kay. In this proposal, the aim is to construct
a $*$-algebra $\AAA(M,g)$ whose elements are interpreted as
polynomials in quantum fields smeared by smooth test functions compactly
supported on a given spacetime $(M,g)$, and such that
$\AAA(M,g)$ possesses the property of {\em F-locality}. This
condition, which will be reviewed in Section~\ref{sect:Alg},
essentially requires that every point in $M$ should have a globally
hyperbolic neighbourhood $N$ such that the {\em induced algebra}
$\AAA(M,g;N)$ (i.e., the subalgebra of $\AAA(M,g)$ consisting
of polynomials of fields smeared by test functions supported
in $N$) should coincide with the {\em intrinsic algebra}
$\AAA(N,g|_{N})$, obtained by regarding $(N,g|_{N})$ as a
spacetime in its own right and following the normal procedure
for globally hyperbolic spacetimes with some choice of time
orientation on $N$. More precise definitions will be given in
Section~\ref{sect:Alg}. If a spacetime admits an F-local $*$-algebra
corresponding to a given field theory, the spacetime is said to be
{\em F-quantum compatible} with this field theory. Kay argues
that non-F-quantum compatible spacetimes would not appear as
classical approximations to states in quantum gravity. The proposal
of Yurtsever~\cite{Yurt} differs from that of Kay in that it
constructs an algebra which generally fails to be F-local, except
in the case of `quantum benign' spacetimes~\cite{Yurt}.

Returning to Kay's proposal, it is clearly important to determine
which spacetimes are
F-quantum compatible with given field theories and which are not.
Three results due to Kay \cite{Kay} shed some light on this issue
for both massive and massless real linear scalar field theory.
Firstly, all globally hyperbolic spacetimes are F-quantum compatible
with these theories;
moreover, the usual minimal algebra on these spacetimes
is F-local. Secondly, {\em any} subspacetime (of zero co-dimension)
of a globally hyperbolic spacetime is also F-quantum compatible with
these field theories. Thirdly, non-time orientable spacetimes are
not F-quantum compatible with either field theory.

As well as these results, Kay discussed two examples of spacetimes
with closed time-like curves. The first, 2-dimensional Misner space
-- in which a region of closed time-like curves `develops' --  was
shown to be non-F-quantum compatible with massless scalar field
theory. The second, the 4-dimensional `spacelike cylinder' (i.e.,
Minkowski space quotiented by a fixed time translation) is F-quantum
compatible with massless real linear scalar field theory. However,
because (in both cases) the assumption of zero mass plays an
essential r\^ole in his argument, Kay left the massive case as an
open question.

In the present paper, we resolve this question for the four
dimensional spacelike cylinder, which we show to be F-quantum
compatible with massive real linear scalar field theory. We also
demonstrate the F-quantum compatibility of the 2-dimensional
spacelike cylinder with both massive and massless
fields.\footnote{The massless case was also known to Kay (private
communication).} Our method in each of these cases is essentially to
find a global bi-solution $\widetilde{\Delta}(x_1;x_2)$ to the
appropriate Klein-Gordon equation which plays the r\^ole of the
advanced-minus-retarded fundamental solution, and which agrees with
the Minkowski space fundamental solution for $x_1$ and $x_2$
sufficiently close together. Once this is found, it is easy to
construct an F-local $*$-algebra for the corresponding real linear
scalar field theory. Our construction is not unique: it turns out that there
are many F-local algebras on the spacetimes we consider. As yet, it
is not clear whether or not these different algebras correspond to
different physics. To resolve this issue, one would have to study the
physically `nice' states (say, quasi-free and Hadamard) on these
algebras, as suggested by Kay \cite{Kay}.

\sect{Algebraic Quantum Field Theory and F-locality}
\label{sect:Alg}

We start by briefly reviewing Kay's proposal \cite{Kay} for the algebraic
approach for real quantum field theory for the Klein-Gordon equation
\begin{equation}
(\Dal_g +m^2)\phi =0
\label{eq:Feq}
\end{equation}
on non-globally hyperbolic, time orientable spacetimes $(M,g)$. This
proposal is developed for the version of algebraic field theory in
which the basic object is a $*$-algebra whose elements are
interpreted as polynomials in smeared quantum fields. (Yurtsever
adopts an alternative viewpoint, in which one considers the Weyl
algebra associated with a suitable symplectic space of classical
solutions -- see, e.g., \cite{KW} -- in his approach to quantisation
on non-globally hyperbolic spacetimes \cite{Yurt}.) In the following,
$\CoinfM$ denotes the space of smooth, real-valued functions
compactly supported on $M$.

We begin by defining a {\em pre-field algebra} to be a $*$-algebra
with identity, $\II$, of polynomials over $\CC$ in abstract objects
$\phi(f)$ labelled by $f\in\CoinfM$ such that the following hold:
\begin{list}{(Q\arabic{enumii})}{\usecounter{enumii}}
\item Hermiticity: $(\phi(f))^*=\phi(f)$ for all $f\in\CoinfM$
\item Linearity: $\phi(\lambda_1f_1+\lambda_2f_2)=\lambda_1\phi(f_1)+
\lambda_2\phi(f_2)$ for all $\lambda_i\in\RR$, $f_i\in\CoinfM$
\item Field Equation: $\phi((\Dal_g+m^2)f)=0$ for all $f\in\CoinfM$.
\end{list}

The motivation for this definition is the interpretation of $\phi(f)$
as the smearing by $f$ of a Hermitian weak solution $\phi(x)$
to~(\ref{eq:Feq}), i.e.,
\begin{equation}
\hbox{`}\phi(f) = \int_M \phi(x) f(x) d\eta \hbox{'},
\end{equation}
where $d\eta=|\det g_{ab}|^{1/2}d^nx$ is the volume element. We emphasise
the heuristic nature of this interpretation: there is no quantum field
$\phi(x)$ as such underlying this approach.

There is a natural mapping between any two pre-field algebras on
$(M,g)$ which identifies elements corresponding to the same
polynomial in the smeared fields. If this mapping is an isomorphism,
we say that the pre-field algebras are {\em naturally isomorphic}.

In general, a pre-field algebra does not represent a quantised field
theory. If $(M,g)$ is globally hyperbolic, there exists a unique
advanced-minus-retarded fundamental solution $\Delta(x_1;x_2)$
to~(\ref{eq:Feq}) (see, e.g., \cite{Lich}) and quantisation is
effected by supplementing relations (Q1--3) with
\begin{list}{(Q4)}{}
\item Covariant Commutation Relations:
$[\phi(f_1),\phi(f_2)] = i\Delta(f_1,f_2)\II$ for all $f_i\in\CoinfM$,
\end{list}
where $\Delta(f_1,f_2)$ is the smeared version of $\Delta(x_1;x_2)$
defined by
\begin{equation}
\Delta(f_1,f_2) = \int_{M\times M} \Delta(x_1;x_2) f(x_1)f(x_2)
d\eta(x_1) d\eta(x_2).
\end{equation}
We will refer to a pre-field algebra satisfying (Q4) as a field
algebra. Such an algebra may be constructed as described
in~\cite{Kay}, by first forming the free $*$-algebra generated (over
$\CC$ and with $\II$) by the abstract objects $\{\phi(f)\mid
f\in\CoinfM\}$ and then quotienting by the relations (Q1--4). We will
refer to the algebra $\AAA(M,g)$ generated in this manner as the
{\em usual field algebra} on $(M,g)$, when $(M,g)$ is globally hyperbolic.
To be precise, the relations (Q1--4) generate a {\em congruence in
algebra}\footnote{A congruence in algebra is a linear equivalence
relation which respects the algebraic operations. Thus in our case,
if the polynomials $P$ and $Q$ are congruent to $P^\prime$ and
$Q^\prime$ respectively, then the product $PQ$ is congruent to
$P^\prime Q^\prime$.} on the algebra of polynomials in the $\phi(f)$
and their adjoints, such that two polynomials are congruent if one
can be manipulated into the form of the other using (Q1--4) and the
usual rules of algebra. The quotient algebra $\AAA(M,g)$ is the
algebra of congruence classes of polynomials. We will abuse notation
by denoting a polynomial and its congruence class in the same way.
For the most part, we will also suppress the mention of congruence
classes.

For general non-globally hyperbolic manifolds $(M,g)$, there is no
advanced-minus-retarded fundamental solution and so one cannot define
a field algebra in the above manner, although there may exist
non-trivial pre-field algebras. Clearly, a criterion is required to
select those pre-field algebras which correspond to some reasonable
notion of quantum field theory. Kay proposed in~\cite{Kay} that this
should be done as follows. Suppose $\AAA(M,g)$ is a pre-field algebra
on a time orientable manifold $(M,g)$ and let $N$ be a globally
hyperbolic subspacetime of $M$.  There are two $*$-algebras naturally
associated with $N$.  Firstly, there is the {\em induced} $*$-algebra
$\AAA(M,g;N)$, namely, the subalgebra of $\AAA(M,g)$ consisting of
polynomials in the $\phi(f_i)$ with all $f_i$ supported in $N$, along
with the identity. $\AAA(M,g;N)$ is easily seen to be a pre-field
algebra on $(N,g|_N)$. Secondly, because $(N,g|_N)$ may be regarded
as a globally hyperbolic spacetime in its own right, there is the
{\em intrinsic} $*$-algebra, $\AAA(N,g|_N)$ which is the usual field
algebra on $(N,g|_N)$. We require that quantisation in each region
$N$ is carried out with respect to a time orientation induced from a
fixed global time orientation on $(M,g)$.\footnote{In~\cite{Kay}, Kay
initially left open the possibility of the time orientation differing
in different regions $N$. However, he then showed that time
orientability was a necessary condition for F-quantum compatibility
of a manifold: we have essentially incorporated this result into our
discussion.} The induced and intrinsic algebras are generally
different (see~\cite{Kay} for explicit examples) because the
commutator in $\AAA(N,g|_N)$ is determined by the
advanced-minus-retarded fundamental solution intrinsic to $(N,g|_N)$
(and therefore depends on the causal structure of $(N,g|_N)$), whilst
that in $\AAA(M,g;N)$ is induced from $\AAA(M,g)$ and may not respect
the causal structure of $(N,g|_N)$. However, Kay's proposal is that,
as a {\em necessary} condition for $\AAA(M,g)$ to be a reasonable
$*$-algebra for quantum field theory on $(M,g)$, these algebras
should agree if $N$ is `sufficiently small'.  More precisely, this is
codified by the notion of F-locality \cite{Kay}:
\begin{Def}[The F-locality Condition]
Let $(M,g)$ be a (not necessarily globally hyperbolic) time
orientable spacetime and suppose $\AAA(M,g)$ is a pre-field algebra
for real linear scalar field theory on $(M,g)$.  $\AAA(M,g)$ is said
to be {\em F-local} if each point $p$ of $M$ has a globally
hyperbolic neighbourhood $N$ for which the induced $\AAA(M,g;N)$ and
intrinsic $\AAA(N,g|_N)$ $*$-algebras are naturally isomorphic. In
addition, $(M,g)$ is said to be {\em F-quantum compatible} if it
admits a pre-field algebra satisfying F-locality.
\end{Def}

In~\cite{Kay}, Kay established various consequences of this
definition, including the result that the usual field algebra on a
globally hyperbolic spacetime is F-local. Thus F-locality provides a
generalisation of the usual quantisation procedure. It is important
to note that, even for F-local algebras, it is not the case that the
induced and intrinsic algebras coincide for {\em any} subspacetime
$N$ -- see \cite{Kay} for discussion.

We now describe the technique we will use to prove the F-quantum
compatibility of the spacelike cylinder spacetimes in
Section~\ref{sect:exam}.  This is accomplished by constructing a
global bi-solution $\widetilde{\Delta}(x_1;x_2)$ to the Klein-Gordon
equation which is equal to the intrinsic advanced-minus-retarded
fundamental solution on sufficiently small globally hyperbolic
subspacetimes (i.e., for $x_1$ and $x_2$ sufficiently close
together). We then use $\widetilde{\Delta}$ to define an F-local
$*$-algebra.

More formally, we define:
\begin{Def} \label{Def:Famr}
Let $(M,g)$ be a (not necessarily globally hyperbolic) time orientable
spacetime, and suppose that $\widetilde{\Delta}(x_1;x_2)$ is an
antisymmetric, real-valued
global bi-solution to the homogenous Klein-Gordon equation on $M\times M$.
We call $\widetilde{\Delta}$ an {\em F-local advanced-minus-retarded
fundamental solution} on $(M,g)$ if every point $p\in M$ has
a globally hyperbolic neighbourhood $N$ such that the restriction
of $\widetilde{\Delta}$ to $N\times N$ agrees with the intrinsic
advanced-minus-retarded fundamental solution on $(N,g|_N)$.
\end{Def}

Here, the property of antisymmetry (i.e.,
$\widetilde{\Delta}(x_1;x_2)=- \widetilde{\Delta}(x_2;x_1)$) is
required because we would like to regard $\widetilde{\Delta}$
(suitably smeared) as $-i$ times the commutator of smeared fields.

With this definition, the following is then almost immediate:
\begin{Thm}
If a spacetime $(M,g)$ admits an F-local advanced-minus-retarded
fundamental solution $\widetilde{\Delta}$ to the Klein-Gordon
equation, then it is F-quantum compatible with real linear scalar
field theory. Moreover, the algebra $\AAA(M,g)$ defined as the
free $*$-algebra generated by the
$\phi(f)$ for $f\in\CoinfM$ quotiented by the relations (Q1)-(Q3) and
\begin{list}{(Q$4^\prime$)}{}
\item Modified CCRs:
$[\phi(f_1),\phi(f_2)] = i\widetilde{\Delta}(f_1,f_2)\II$
for all $f_i\in\CoinfM$
\end{list}
is F-local, where $\widetilde{\Delta}(f_1,f_2)$ is the smeared
version of $\widetilde{\Delta}(x_1;x_2)$.
\end{Thm}
{\em Proof:} Let $p$ be an arbitrary point in $M$ and let $N$ be
a globally hyperbolic neighbourhood of the type guaranteed by
Definition~\ref{Def:Famr}.
Then the F-locality of $\widetilde{\Delta}$ implies that the natural
mapping between $\AAA(N,g|_N)$ and $\AAA(M,g;N)$ is an isomorphism,
because the relations (Q1--3) and (Q$4^\prime$) for $(M,g)$ agree with
the relations (Q1--4) for $(N,g|_N)$ on polynomials involving only
test functions supported in $N$.~$\square$

Thus it suffices to exhibit an F-local advanced-minus-retarded
fundamental solution on a spacetime $(M,g)$ in order to conclude its
F-quantum compatibility and to construct an F-local algebra there.

In addition to F-locality, we will require that our $*$-algebra $\AAA(M,g)$
should be globally covariant with respect to the isometries of the
spacetime (cf.~\cite{Dimock}). That is, we require there to be
a representation $\kappa\mapsto\alpha_\kappa$ of the isometry
group of $(M,g)$ as automorphisms on $\AAA(M,g)$ such that
$\alpha_\kappa(\phi(f)) = \phi(f\circ\kappa)$ for each $f\in\CoinfM$
and isometry $\kappa$.
This places constraints on the F-local advanced-minus-retarded
fundamental solution $\widetilde{\Delta}(x_1;x_2)$, because we have
\begin{equation}
i\widetilde{\Delta}(f\circ\kappa,g\circ\kappa)\II = \alpha_\kappa(
i\widetilde{\Delta}(f,g)\II)
\end{equation}
for all $f,g\in\CoinfM$. Thus we require $
\widetilde{\Delta}(\kappa(x_1);\kappa(x_2))= \pm
\widetilde{\Delta}(x_1;x_2)$ for each isometry $\kappa$, with the
sign depending on whether
$\alpha_\kappa$ is a linear ($+$) or anti-linear ($-$) automorphism.

\sect{Quantum Field Theory on Spacelike Cylinders}
\label{sect:exam}

In this section, we discuss the F-quantum compatibility of the 2- and
4-dimensional spacelike cylinders, which are obtained as quotients of
2- or 4-dimensional flat Minkowski space by a fixed time-like
translation. We write coordinates for these
spacetimes in the form $(t,\xvec)$ and, in the 2-dimensional case,
write the single spatial coordinate as $z$.
The spacelike cylinder spacetime is then defined to be the quotient
of Minkowski space by the fixed time translation $t\rightarrow t+T$
for some $T>0$.

In the following, it will be useful to
have an explicit class of globally hyperbolic neighbourhoods
available: accordingly, we define a {\em diamond of size $d$} about
position $(t,\xvec)$ to be the region
\begin{equation}
N_d=\{(t^\prime,\xvec^\prime)\mid~
|t^\prime-t|+|\xvec^\prime-\xvec|<d \}.
\end{equation}
In our example spacetimes every point has a globally
hyperbolic diamond neighbourhood of any size less than $T/2$. If a
diamond neighbourhood in one of the spacelike cylinders is
globally hyperbolic, its intrinsic advanced-minus-retarded fundamental
solution is just the restriction of that for 2- or 4-dimensional
Minkowski space as appropriate.

\subsection{The 2-dimensional Spacelike Cylinder}
\label{sect:2dsc}

We now show that the 2-dimensional spacelike cylinder spacetime is
F-quantum compatible with both massless and massive real linear
scalar field theory, by exhibiting suitable F-local
advanced-minus-retarded fundamental solutions.

As mentioned above, we will require global covariance with respect to
the isometries of the spacetime: these are generated by the group of
spacetime
translations and the two discrete symmetries of parity and time
reversal. The translations form a continuous group of isometries and
must therefore be implemented by linear automorphisms at the
algebraic level, whilst parity and time reversal are implemented as
usual by linear and anti-linear automorphisms respectively (locally,
this is actually a consequence of F-locality). In consequence,
$\widetilde{\Delta}(x_1;x_2)$ can be written as a function of
$x_1-x_2$ alone (by translational invariance):
$\widetilde{\Delta}(x_1;x_2) = \widetilde{\Delta}(t_1-t_2,z_1-z_2)$,
where $\widetilde{\Delta}(t,z)$ is even in $z$ and odd in $t$.

We now prove the F-quantum compatibility of the 2-dimensional
spacelike cylinder, treating the massless and massive cases
separately.

{\noindent\bf Massless Klein-Gordon Theory}

It is helpful to begin by recalling the
definition of the advanced-minus-retarded fundamental solution to
the massless Klein-Gordon equation on 2-dimensional Minkowski space.
This is given by $\Delta(t_1,z_1;t_2,z_2)=\Delta(t_1-t_2,z_1-z_2)$,
where $\Delta(t,z)$ solves
\begin{equation}
\left(\frac{\partial^2}{\partial t^2} -
\frac{\partial^2}{\partial z^2} \right) \Delta(t,z) = 0
\label{eq:KG2}
\end{equation}
with initial data
\begin{equation}
\Delta(0,z) = 0, \qquad \left.\frac{\partial\Delta}{\partial t}
(t,z)\right|_{t=0}=-\delta(z).
\end{equation}
As is well known, the solution is
\begin{equation}
\Delta(t,z) = -\frac{1}{2} \vep(t)\vth(t^2-z^2),
\label{eq:amrf1}
\end{equation}
where $\vth(x)$ is the Heaviside function, and $\vep(x)$ is the sign
of $x$.  Thus $\Delta(t,z)$ is equal to $-\frac{1}{2}$ in the future
light-cone of the origin, $\frac{1}{2}$ in the past light-cone and
vanishes elsewhere.

To construct an F-local advanced-minus-retarded fundamental solution
on the spacelike cylinder, we regard it as the strip
$|t|<T/2$ in Minkowski space and impose periodic boundary conditions
on all fields at $\pm T/2$. We seek a solution $\widetilde{\Delta}$
to~(\ref{eq:KG2}) which is periodic in time with period $T$ and
which agrees with $\Delta(t,z)$ for $t,z$ sufficiently close to the
origin. There are many possibilities for $\widetilde{\Delta}$
(see the corresponding discussion for the massive case); we
give the simplest, which is
\begin{equation}
\widetilde{\Delta}(t,z) = \sum_{n=-\infty}^\infty
(-1)^n  \Delta(t, z-nT/2).
\end{equation}
The sum converges (because only finitely many summands are non-zero
at any given $(t,z)$) and its values
are displayed in Figure 1.
Note that $\widetilde{\Delta}$ is
periodic in both $t$ and $z$ with period $T$, so it suffices to
show the values for $|t|,|z|<T/2$. $\widetilde{\Delta}$ is clearly
the solution to~(\ref{eq:KG2}) with initial data
\begin{equation}
\widetilde{\Delta}(0,z) = 0,
\qquad \left.\frac{\partial\widetilde{\Delta}}{\partial t}
(t,z)\right|_{t=0}=\sum_{n=-\infty}^\infty (-1)^{n+1}\delta(z-nT/2).
\end{equation}
One may also regard $\widetilde{\Delta}(t,z)$ as the result
of `wrapping' the Minkowski space fundamental solution $\Delta$ round
a spacelike cylinder of period $T/2$ with {\em anti-periodic}
boundary conditions. It is interesting to compare this with Kay's proof
of F-quantum
compatibility for massless fields on the 4-dimensional spacelike
cylinder, which proceeds by wrapping the Minkowski fundamental
solution round the cylinder of period $T$ with periodic boundary
conditions.

Finally, it is easy to see from Fig. 1 that
$\widetilde{\Delta}(t_1,z_1;t_2,z_2)
=
\widetilde{\Delta}(t_1-t_2,z_1-z_2)$ agrees with
$\Delta(t_1,z_1;t_2,z_2)$ within any diamond neighbourhood of size
less than $T/4$ about any point $p$ in the spacetime. (Note that
if two points lie in a diamond region of size $d$ of some point,
then their difference lies in a diamond region of size $2d$ about
the origin.) Thus $\widetilde{\Delta}$ is an F-local
advanced-minus-retarded fundamental solution on the two dimensional
spacelike cylinder.
As an immediate corollary, due to the periodicity of $\widetilde{\Delta}$
in both $t$ and $z$, we also conclude F-quantum compatibility of
the 2-torus with equal periods with massless field theory.

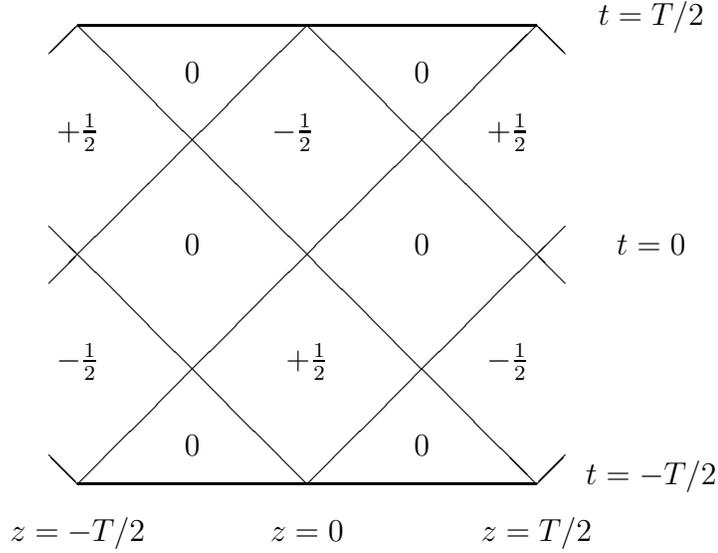
\begin{figure}
\setlength{\unitlength}{0.0075in}%
\center{\begin{picture}(440,378)(120,320)
\thicklines
\put(160,680){\line( 1, 0){320}}
\put(160,360){\line( 1, 0){320}}
\thinlines
\put(160,360){\line( 1, 1){320}}
\put(160,680){\line( 1,-1){320}}
\put(320,360){\line( 1, 1){180}}
\put(320,360){\line(-1, 1){180}}
\put(320,680){\line( 1,-1){180}}
\put(320,680){\line(-1,-1){180}}
\put(160,360){\line(-1, 1){20}}
\put(160,680){\line(-1,-1){20}}
\put(480,360){\line( 1, 1){20}}
\put(480,680){\line( 1,-1){20}}

\put(310,600){\makebox(0,0)[c]{\raisebox{0pt}[0pt][0pt]{$-\frac{1}{2}$}}}
\put(240,520){\makebox(0,0)[c]{\raisebox{0pt}[0pt][0pt]{$0$}}}
\put(400,520){\makebox(0,0)[c]{\raisebox{0pt}[0pt][0pt]{$0$}}}
\put(460,600){\makebox(0,0)[c]{\raisebox{0pt}[0pt][0pt]{$+\frac{1}{2}$}}}
\put(240,640){\makebox(0,0)[c]{\raisebox{0pt}[0pt][0pt]{$0$}}}
\put(400,640){\makebox(0,0)[c]{\raisebox{0pt}[0pt][0pt]{$0$}}}
\put(240,380){\makebox(0,0)[c]{\raisebox{0pt}[0pt][0pt]{$0$}}}
\put(400,380){\makebox(0,0)[c]{\raisebox{0pt}[0pt][0pt]{$0$}}}
\put(160,600){\makebox(0,0)[c]{\raisebox{0pt}[0pt][0pt]{$+\frac{1}{2}$}}}
\put(160,440){\makebox(0,0)[c]{\raisebox{0pt}[0pt][0pt]{$-\frac{1}{2}$}}}
\put(320,440){\makebox(0,0)[c]{\raisebox{0pt}[0pt][0pt]{$+\frac{1}{2}$}}}
\put(460,440){\makebox(0,0)[c]{\raisebox{0pt}[0pt][0pt]{$-\frac{1}{2}$}}}
\put(320,320){\makebox(0,0)[c]{\raisebox{0pt}[0pt][0pt]{$z=0$}}}
\put(480,320){\makebox(0,0)[c]{\raisebox{0pt}[0pt][0pt]{$z=T/2$}}}
\put(160,320){\makebox(0,0)[c]{\raisebox{0pt}[0pt][0pt]{$z=-T/2$}}}
\put(560,520){\makebox(0,0)[c]{\raisebox{0pt}[0pt][0pt]{$t=0$}}}
\put(560,360){\makebox(0,0)[c]{\raisebox{0pt}[0pt][0pt]{$t=-T/2$}}}
\put(560,680){\makebox(0,0)[c]{\raisebox{0pt}[0pt][0pt]{$t=T/2$}}}
\end{picture}}
\caption{The values taken by the function $\widetilde{\Delta}(t,z)$ on the
spacelike cylinder.}
\end{figure}

\vspace{0.2truecm}
{\noindent\bf Massive Klein-Gordon Theory}

In this case, we seek a global solution $\widetilde{\Delta}(t,z)$ to
the massive Klein-Gordon equation on the spacelike cylinder, which
agrees with the usual advanced-minus-retarded fundamental solution on
2-dimensional Minkowski space
\begin{equation}
\Delta(t,z) = -\frac{1}{2}\vep(t)\vth(t^2-z^2)J_0(m(t^2-z^2)^{1/2})
\end{equation}
for $t$ and $z$ sufficiently close to the origin. Here, $J_0(z)$ is
the Bessel function of order zero. As mentioned above, we will
require $\widetilde{\Delta}(t,z)$ to be even in $z$ and odd in $t$.

The key to constructing such a function is that the initial
value problem {\em is}
well-posed for the Klein-Gordon equation
on the spacelike cylinder, when one specifies initial data
\begin{equation}
f(t) = \phi(t,0) \quad {\rm and}\quad g(t) = \left.
\frac{\partial\phi}{\partial z}(t,z)\right|_{z=0}
\label{eq:IVP}
\end{equation}
on the circle $z=0$, where $f(t)$ and $g(t)$ are periodic with
period $T$. (Essentially, we are just turning the spacelike
cylinder on its end to obtain the globally hyperbolic `timelike
cylinder', with $t$ regarded as a periodically identified
`spatial' coordinate and $z$ regarded as `time'.)

We now set $g(t)$ to be identically zero (reflecting the fact that we
want $\widetilde{\Delta}(t,z)$ to be even in $z$), and require that
$f(t)$ should coincide with
$\Delta(t,0)=-\frac{1}{2}\vep(t)J_0(m|t|)$ for $|t|<T/2-\epsilon$ for
some $\epsilon>0$. Outside this region, we can choose $f(t)$
arbitrarily, subject to the requirements that it be real-valued, odd,
smooth and obey periodic boundary conditions at $t=\pm T/2$. From
elementary properties of hyperbolic equations, it follows that the
solution $\widetilde{\Delta}(t,z)$ to the massive Klein-Gordon
equation with this initial data agrees with $\Delta(t,z)$ in the
diamond of size $T/2-\epsilon$ about the origin.  Thus
$\widetilde{\Delta}(x_1;x_2)= \widetilde{\Delta}(t_1-t_2,z_1-z_2)$ is
an F-local advanced-minus-retarded fundamental solution for the
massive Klein-Gordon equation on the spacelike cylinder.

There is clearly a substantial non-uniqueness in our construction,
because $f(t)$ can be chosen freely (amongst smooth, odd functions)
for $T/2-\epsilon<|t|<T/2$ (at $t=T/2$, $f$ must vanish, along
with all its even derivatives).  Thus there are very many F-local
algebras on this spacetime.  However, in general, the corresponding
$\widetilde{\Delta}(t,z)$ functions are not bounded as
$|z|\rightarrow\infty$ for each $t$, owing to the presence of
exponentially growing modes in the solution to the initial value
problem~(\ref{eq:IVP}).\footnote{This is the penalty for switching
the interpretations of $t$ and $z$: the `mass' becomes imaginary!}

To see this explicitly, note that mode solutions to
the massive Klein-Gordon equation
\begin{equation}
\left(\frac{\partial^2}{\partial t^2} -
\frac{\partial^2}{\partial z^2} +m^2\right) \phi(t,z) = 0
\label{eq:KG2m}
\end{equation}
with periodic time dependence $e^{ 2\pi int/T}$ are
$\phi_{n,\pm}(t,z)\propto e^{2\pi int/T \pm i k_n z}$, where
$k_n$ is determined by
\begin{equation}
k_n = \left( \frac{4\pi^2n^2}{T^2} - m^2 \right)^{1/2}.
\end{equation}
Accordingly, for any $m>0$ there is at least one mode (solution
with imaginary $k_n$) which grows exponentially as $|z|\rightarrow\infty$.

Presumably, one should exclude solutions with unbounded
$\widetilde{\Delta}(t,z)$. One can arrange this by using the freedom
in $f(t)$ to
ensure that it possesses no harmonics corresponding to growing modes.
Defining
\begin{equation}
\ell_n(f) = \int_{-T/2}^{T/2} e^{2\pi i nt/T} f(t) dt
\end{equation}
for $n\in\ZZ$, this is equivalent to requiring $\ell_n(f) = 0$
for $|n|\le mT/2\pi$. We note that the $n=0$ harmonic vanishes in any
case because $f(t)$ is odd; thus for $mT<2\pi$ (small cylinders or
light particles) growing modes are automatically absent.

More generally, because $f(t)$ is chosen to be real-valued and odd,
we have
$\ell_{-n}(f) = -\ell_n(f)=\ell_n(f)^*$, and so growing modes are
excluded
provided that $\ell_{n}(f) = 0$ for $1\le n\le mT/2\pi$. This
can be arranged by choosing $f(t)$ to have the form
\begin{equation}
f(t) = -\frac{1}{2}\vep(t)J_0(m|t|)\chi(t) +
\sum_{1\le k \le mT/2\pi} \lambda_k h_k(t),
\end{equation}
with $\lambda_k\in\RR$,
where $\chi(t)$ is smooth, even, compactly supported in
$|t|<T/2-\epsilon/2$, and equal to unity for $|t|<T/2-\epsilon$, and
the $h_k(t)$ are smooth odd functions compactly supported in
$T/2-\epsilon/2<|t|<T/2$ such that $\ell_j(h_k)=0$ for
$0\le j<k$ and $\ell_k(h_k)\not =0$. The functions $\chi$ and $h_k$
are also required to be real-valued.

To construct suitable $h_k$, we define the function spaces
\begin{equation}
\DD_n = \left\{ \varphi\in\DD_0 \mid \ell_j(\varphi)
=0~{\rm for}~j=1,2,\ldots,k\right\}
\end{equation}
where $\DD_0$ is the space of smooth odd real-valued functions
compactly supported
in $T/2-\epsilon/2<|t|<T/2$. For each $k$, $h_k$ is chosen to be a
representative of any non-zero equivalence class in the quotient
$\DD_{k-1}/\DD_k$, which is non-trivial because $\ell_1,\ldots,\ell_k$
are linearly independent distributions on $\DD_0$. The $h_k$ clearly
possess the required properties and the $\lambda_k$ may therefore
be chosen to ensure that $\ell_n(f)=0$ for each $|n|\le mT/2\pi$ as
required. Reality of the $\lambda_k$ follows from the fact that
$\ell_n(\varphi)$ is purely imaginary for odd real-valued $\varphi$.

One could, of course, specify a distinguished $\widetilde{\Delta}$
by letting $\epsilon$ shrink to zero, thereby putting all the
ambiguity in $f(t)$ into a distributional contribution supported at
the point $t=\pm T/2$. The distinguished $\widetilde{\Delta}$ would
be obtained by requiring this distribution to be as `regular' as
possible, but nonetheless singular enough to ensure the absence of
growing modes. In fact, the F-local advanced-minus-retarded
solution identified in the massless
case (where there are in fact no growing modes) can be regarded as a
(trivial) example of this procedure.

\subsection{The 4-dimensional Spacelike Cylinder}

As we have mentioned, it has already been shown by Kay \cite{Kay}
that the 4-dimensional spacelike cylinder is F-quantum compatible
with massless scalar fields. Here, we demonstrate that the same is
true for the massive case.

The isometries of the 4-dimensional spacelike cylinder are generated
by the group of spacetime translations, the group of rotations, and
the discrete symmetries of parity and time reversal. Of these
generators, only time reversal is represented by an anti-linear
automorphism.  The requirement of covariance with respect to these
isometries therefore entails that $\widetilde{\Delta}(x_1;x_2)=
\widetilde{\Delta}(t_1-t_2,|\xvec_1-\xvec_2|)$, where
$\widetilde{\Delta}(t,r)$ is odd in $t$.

In addition, we require that $\widetilde{\Delta}(x_1;x_2)$ should
agree with the advanced-minus-retarded fundamental solution for
4-dimensional Minkowski space given by $\Delta(x_1;x_2) =
\Delta(t_1-t_2,|\xvec_1-\xvec_2|)$, where
\begin{equation}
\Delta(t,r) =m\vep(t)\vth(t^2-r^2)
\frac{J_1(m(t^2-r^2)^{1/2})}{4\pi(t^2-r^2)^{1/2}} -
\frac{\vep(t)}{2\pi}
\delta(t^2-r^{2}).
\end{equation}

The essential point here is that spherical symmetry has reduced
this to an effectively 2-dimensional problem. Writing
$\widetilde{\Delta}(t,r)=u(t,r)/r$,
$u(t,r)$ satisfies
\begin{equation}
\left(\frac{\partial^2}{\partial t^2} -
\frac{\partial^2}{\partial r^2} +m^2\right) u(t,r) = 0,
\label{eq:KG4m}
\end{equation}
and we may proceed much as in Section~\ref{sect:2dsc}, by specifying
initial data on the circle $r=0$, defined by
\begin{equation}
f(t) = u(t,0) \quad{\rm and}\quad
g(t) = \left.
\frac{\partial u}{\partial r}(t,r)\right|_{r=0} .
\end{equation}

We impose data with $f(t)$ identically zero and $g(t)$ given by
\begin{equation}
g(t) = m\vep(t)\frac{J_1(mt)}{4\pi t} + \frac{\delta^\prime(t)}{2\pi}
\quad {\rm for}\quad |t|<T/2-\epsilon
\end{equation}
for some $\epsilon>0$. For $t$ outside this region, $g(t)$ is chosen
to be smooth, odd and to obey periodic boundary conditions at
$t=\pm T/2$. We also use some of this freedom to ensure the absence of
tachyonic modes in the same manner as before.

These data agree with the corresponding quantities obtained from
the flat spacetime fundamental solution $\Delta(t,r)$ for
$|t|<T/2-\epsilon$ as may be seen by a simple distributional
calculation, which rests on the following easily established result:
Let $\psi_r(t)=r\varepsilon(t)\delta(t^2-r^2)$
be a family of distributions labelled by $r$. Then
\begin{list}{(\alph{enumii})}{\usecounter{enumii}}
\item  $\psi_r(t) = \frac{1}{2}(\delta(t-r)-\delta(t+r))
\rightarrow 0$ as $r\rightarrow 0^+$
\item  $(\partial \psi_r/\partial r)(t) =
-\frac{1}{2}(\delta^\prime(t-r)+\delta^\prime(t+r)) \rightarrow -
\delta^\prime(t)$ as $r\rightarrow 0^+$.
\end{list}

Thus, the solution $\widetilde{\Delta}(t,r)$ obtained from this
initial data agrees with $\Delta(t,r)$ for $|t|+r<T/2-\epsilon$. Thus
$\widetilde{\Delta}(x_1;x_2)=
\widetilde{\Delta}(t_1-t_2,|\xvec_1-\xvec_2|)$
agrees with $\Delta(x_1;x_2)$ when $x_1,x_2$ lie in a diamond
neighbourhood of size $T/4-\epsilon/2$ of any given point, so
$\widetilde{\Delta}$ is an F-local advanced-minus-retarded
fundamental solution to the massive Klein-Gordon equation on the
4-dimensional spacelike cylinder. This spacetime is therefore
F-quantum compatible with massive real linear scalar field theory.
As in the 2-dimensional case, our construction is highly non-unique:
similar comments to those made in that case also apply here.
We also note that by setting $m=0$ in the above, we obtain a new
proof of the F-quantum compatibility of this spacetime with
massless field theory.

\section{Conclusion}

We have studied the 2- and 4-dimensional spacelike cylinders and have
shown that the 2-dimensional cylinder is F-quantum compatible with
both massive and massless fields, and that the 4-dimensional cylinder
is F-quantum compatible with massive fields. This complements the
earlier result of Kay on the F-quantum compatibility of the
4-dimensional spacelike cylinder with massless fields and resolves
the question raised in~\cite{Kay} as to whether or not this
spacetime is F-quantum compatible with massive fields. These results
help to strengthen the claim that F-locality provides a good basis
for quantum field theory on non-globally hyperbolic spacetimes.

Our construction exhibits a considerable degree of non-uniqueness. A
similar phenomenon was noted by Kay for spacetimes which can be
isometrically embedded in a globally hyperbolic spacetime: here we
have shown that this can happen even for spacetimes which cannot be
embedded in this way. Non-uniqueness might present a serious problem
if it transpired that different F-local algebras could correspond to
different physics. To investigate this, it would be necessary
to move beyond the construction of the algebra of observables, to the
study of physically reasonable states on the algebra. If necessary,
it might be possible to remove this non-uniqueness by imposing further
conditions on the algebra of observables, or on the class of allowed
states, as mentioned in~\cite{Kay}.

Finally, our treatment of the spacelike cylinders relied heavily on
the special property that they may be turned on end to yield globally
hyperbolic spacetimes. More generally, it is of interest to determine
what conditions a spacetime must satisfy in order to admit an F-local
advanced-minus-retarded fundamental solution. This appears to be
closely related to whether or not a spacetime is `classically
benign'~\cite{Yurt2} with respect to weak solutions of the
Klein-Gordon equation. We note that (almost by definition) a
translationally invariant classically benign spacetime admits an
F-local advanced-minus-retarded fundamental solution, and is
therefore F-quantum compatible. More generally, one needs to
generalise the notion of classical benignity to treat bi-solutions.
Progress in the classification of such manifolds would be of
considerable interest in the context of F-locality.

{\noindent\bf Acknowledgments:} We are grateful to Bernard Kay, Petr
H\'aj\'{\i}{\v c}ek and Peter Minkowski for useful conversations and
comments. CJF thanks the Royal Society and we both thank the
Schweizerischer Nationalfonds for financial support.

\end{document}